# A Guided Inquiry Approach to Students Co-Designing Generative AI Course Policies

Ashish Hingle
Orfalea College of Business
Cal Poly, San Luis Obispo, CA, USA
ahingle@calpoly.edu

Aditya Johri
College of Engineering & Computing
George Mason University, Faifrax, VA, USA
johri@gmu.edu

**Abstract**

As generative AI (GenAI) use among students increases, educators face growing questions about how to support learning while addressing ethical and institutional concerns. This exploratory study examines a guided inquiry activity in which students co-designed a GenAI course policy. Students first developed individual policy proposals focused on appropriate and ethical use of GenAI, then collaboratively refined them by incorporating diverse stakeholder perspectives. The following research questions guided the study: 1) what practical factors do students prioritize in their GenAI use policies, and how do they justify these choices? and 2) how do participants reflect on the policy design process? Participants first completed readings, then used GenAI to brainstorm initial policy ideas. Next, they articulated their own perspectives through a written assignment and a course policy they designed individually. Finally, they incorporated diverse stakeholder perspectives by collaborating with peers to develop a collective policy. Analysis of student artifacts and group discussions showed that participants prioritized training for students and instructors, standardized procedures for disclosing AI use, and stronger institutional support. Participants also wanted greater involvement in GenAI-related decision-making. They described the policy design process as a way to engage with multiple perspectives and the inherent trade-offs involved in governing AI use. This study offers pedagogical insights into how policy co-design activities can surface student values, concerns, and sensemaking about GenAI in educational contexts.

## I. Introduction

The use of artificial intelligence (AI) and the adoption of generative AI (GenAI) in the classroom have been topics of ongoing discussion, with many institutions grappling with the acceptable boundaries for its implementation (Ali et al., 2025). As a result, GenAI tools are gaining traction in academic settings, with some institutions adopting more formalized policies governing their use. However, these policies are not yet universally consistent, as varied approaches and enforcement mechanisms have emerged across different institutions in the US (McDonald et al., 2025). Furthermore, as many educational institutions seek to acquire institutional licenses for GenAI tools, developing standardized course policies that align curricula with learning outcomes has become increasingly important (Wang et al., 2024). Unfortunately, this process remains inconsistent. While there is room to discuss the extent to which students' perceptions of acceptable use should affect policy, students are largely absent from the policy-making process altogether.

Designing without input raises concerns, as research shows that students are actively engaged with questions about what AI is and how to use it in their learning (Hingle et al., 2022; Johri et al., 2024).

Recognizing the importance of incorporating students' voices in discussions about AI in education, this study examines how students conceptualize and construct course policies governing the use of GenAI in academic settings. Participants were enrolled in a course examining AI's impact across disciplines and skills, giving them relevant grounding for this policy design work. We present a novel learning activity in which students actively design a course policy for generative AI, deepening their understanding of AI concepts while experiencing a multi-stakeholder policy design and decision-making process firsthand. We used a guided inquiry process to provide support and allow participants to independently conduct research to understand a topic and navigate their perspectives with their peers. The purpose of this study is twofold: to present details on the activity we designed with students and make these available for course instructors and administrators to use, and to present the empirical findings from our implementation of the activity.

The research questions guiding this work are: 1) What practical factors do students prioritize in their proposed GenAI use policies, and how do they justify these choices?; and 2) How do participants reflect on the process of designing a GenAI policy?

## II. Literature Review and Conceptual Model

While existing AI literacy interventions primarily emphasize competencies, awareness, or ethical understanding, this work conceptualizes the GenAI policy itself as both an artifact for student engagement and a pedagogical tool. By engaging students in the co-design of course policies, the study reframes policy not as a fixed institutional document but as a social process through which students negotiate acceptable use, responsibility, and academic values. This bottom-up perspective complements existing top-down analyses of generative AI policy in higher education by centering student agency in policy formation. The following review situates this contribution within existing scholarship.

### A. Conceptual Model

Figure 1 presents the conceptual model for this study. Addressing a gap in AI literacy research, the model positions students not only as learners but also as co-designers of course structures and policies that affect them. Guided inquiry frames the activity, while perspectival thinking, informed by situated learning theory and principles of student voice, helps students examine AI policy guidelines from multiple stakeholder perspectives. By integrating these theoretical framings with AI literacy categories, the model emphasizes a participatory design process grounded in student engagement. We briefly discuss these elements below.

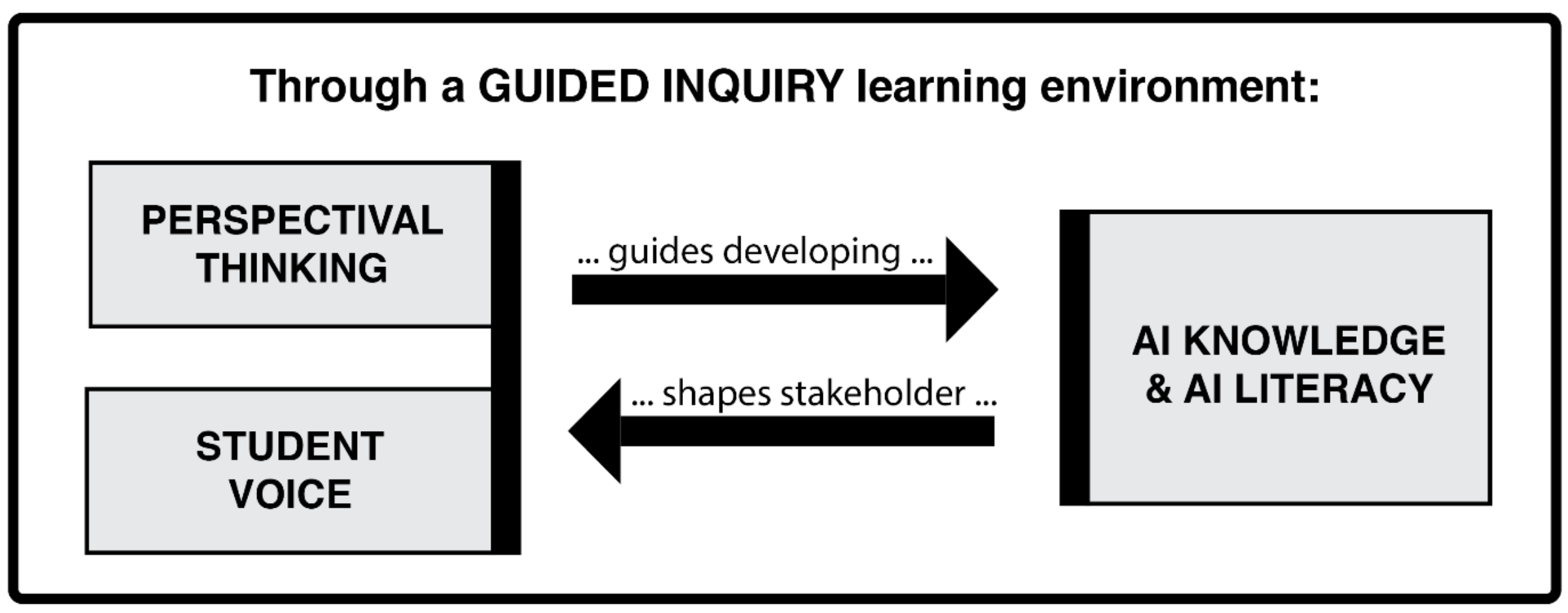

*Figure 1: The conceptual model guiding this study. Through a guided inquiry learning environment, perspectival thinking, and student voice guide the development of AI knowledge and AI literacy. This, in turn, helps shape stakeholder perspectival thinking and student voice.*

B. Guided Inquiry Approach

Guided inquiry, a subset of inquiry-based learning, prepares students for lifelong learning by fostering deep thinking and connections between learners, content, and real-world applications (Kuhlthau et al., 2007). Guided inquiry integrates inquiry-based assessments into the learning process "based on the conviction that science learning is more than the memorisation of scientific facts and information, but rather is about understanding and applying scientific concepts and methods" (Bell et al., 2010, p. 350). Through a guided learning approach, students actively engage in the content through thoughtful planning and adaptability (Todd et al., 2005). The instructional team works together, bringing their combined expertise to connect the curriculum to students' worlds and provide a comprehensive, thought-provoking learning experience (Kuhlthau et al., 2007, p. 5). Rooted in constructivism, inquiry-based curricula provide opportunities for students to explore authentic scientific phenomena, participate in generating research questions, conduct investigations, draw their own conclusions, and communicate their findings to peers (Tseng et al., 2013). It also integrates transferable information literacy skills, helping students locate, evaluate, and use information effectively throughout the research process (Kuhlthau et al., 2007, p. 79).

In the context of GenAI, the use of inquiry-learning design remains relatively underexplored. Inquiry-learning provides a pedagogical framework that aligns well with efforts to support the development of AI literacy, as it emphasizes student-centered exploration, critical questioning, and scaffolded problem-solving (Tang et al., 2026). Within AI education, guided inquiry enables learners to investigate how these systems operate and their ethical and societal implications, actively constructing understanding rather than passively receiving information. This approach cultivates not only technical comprehension but also reflective and ethical reasoning, empowering students to become informed, critical participants in AI-mediated environments.

C. Engaging Perspectival Thinking and Student Voice

While the guided inquiry approach provides students with a structure for engaging with questions and content, perspectival thinking, derived from situated learning, enables learners to consider multiple stakeholders when thinking through a situation such as the design of course policies. Greeno and van de Sande's work within the situated learning tradition offers a perspective on students' learning that encourages them to engage with different viewpoints and stakeholders. They argue that an individual's or group's understanding of any concept is rooted in their ability to construct perspectival understandings grounded in specific activities and governed by underlying principles (Greeno & van de Sande, 2007). Perspectival understanding incorporates entities, their properties, and relationships in creating knowledge and understanding. The critical aspect of this process is that the learner brings their point of view to the situation (Hingle & Johri, 2024b). In this sense, learning can occur when students construct coherent understandings from perspectives both within and outside their own experience.

Greeno and van de Sande's framework posits that perspectives function at various levels, but they primarily focus on information perspectives, namely the situation being addressed and the informational resources associated with it that shape participants' understanding (Greeno & van de Sande, 2007). This notion of perspectival understanding aligns with how we conceptualize the learning experience through the GenAI module. To grasp how algorithms function in real-world contexts, a situated learning theory-based approach can be employed to emphasize the significance of context and multiple perspectives in designing effective pedagogical tools (Lave & Wenger, 1991). Through a situated lens, students should be provided with both authority, namely the power or right to act on behalf of others, and accountability, being answerable for one's actions and decisions. Together, these elements create opportunities for all participants to make decisions and contribute to the learning experience collaboratively (Greeno, 2011).

This focus on authority, accountability, and multiple perspectives aligns with participatory and co-design approaches, which emphasize designing with stakeholders rather than for them (Sanders & Stappers, 2008). In educational contexts, co-design positions learners as contributors to the learning environments, tools, and policies that shape their experiences. Here, co-design does not imply that instructors or students alone bear responsibility for GenAI policy development; rather, each perspective can meaningfully inform it. This framing supports the GenAI policy design activity as a means for students to articulate values, negotiate trade-offs, and contribute to decisions that affect their learning.

When considering key stakeholder perspectives in co-design processes, students offer a vital and often underrepresented viewpoint (Bourke & MacDonald, 2018; Hingle & Johri, 2024a). As the primary group directly affected by such policies, student insights are essential to ensuring an equitable, transparent, and supportive learning environment. Research on student voice consistently highlights the importance of involving students as genuine stakeholders rather than passive recipients of policy decisions (Hart, 1992; Fielding, 2004; Cook-Sather, 2006). Fielding (2004) notes that authentic participation extends beyond consultation to partnership and co-construction, where both students and teachers are engaged as central perspectives (p. 306-307). Likewise, Rudduck & Fielding (2006) emphasize that meaningful involvement fosters a sense of belonging and agency; not simply being asked for an opinion but also finding the words through their experiences to express what they want to say (p. 224).

In AI policy development, situated learning and student voice encourage students to consider multiple stakeholder perspectives while reflecting on their own experiences as primary stakeholders (Hingle et al., 2022). This dual focus aligns with AI literacy efforts that promote critical, context-aware, and ethically informed engagement with AI.

### D. AI Literacy Framework

Finally, recent work on AI literacy underpins the learning objectives and the primary framework for data analysis. Existing reviews of AI literacy emphasize the importance of context-specific approaches that align with learners' educational levels, disciplinary backgrounds, and goals for engaging with AI. These goals range from developing technical proficiency to fostering critical understanding and ethical awareness (Chiu, 2025; Long & Magerko, 2020; Ng et al., 2021). The reviews highlight that AI literacy in practice is often supported through different levels of learning, progressing along a hierarchy similar to Bloom's taxonomy, from recognizing or understanding basic AI functionality to creating with AI. Generally, studies exploring AI literacy across educational levels organize it into four components: *understanding*, *applying*, *evaluating*, and *creating AI* (Ng et al., 2021; Casal-Otero et al., 2023; Almatrafi et al., 2024).

Developing AI literacy activities that account for learners' contexts remains an important and ongoing area of work, particularly as educators seek to integrate constructs such as perspectival thinking and student voice, both central to this study, into the design of learning experiences. To connect these ideas with existing models of AI education, we draw on the *ED-AI Lit framework* (Allen & Kendeou, 2024) to situate perspectival thinking and student voice within a structured approach to AI literacy. Together, these elements provide a foundation for understanding how learners critically and ethically engage with AI (Allen & Kendeou, 2024). Specifically, the elements are:

I. **Knowledge**: to gain an understanding of how AI technologies work and their underlying principles;
II. **Evaluation**: to develop the ability to critically judge AI technologies, considering their strengths, limitations, and potential biases;
III. **Collaboration**: to develop skills for effective communication and collaboration with AI systems and individuals;
IV. **Contextualization**: to understand how to use AI as a tool in real-world settings;
V. **Autonomy**: to develop self-determination in actions and decision-making when interacting with AI; and
VI. **Ethics**: to recognize and address moral issues related to AI technologies, including fairness, accountability, transparency, and privacy.

## III. Study Context

### A. Generative AI Module

We conducted this study in an undergraduate course, “Technology and the Global Society,” which fulfills the technology ethics requirement in a U.S. College of Engineering and Computing. The course examines the role of information technology in societal and systemic contexts. The course is designed using active learning to scaffold key concepts while encouraging students to extend their learning through independent research and engagement with news, academic literature, and industry reports.

The course included four 3-4-week modules over the semester, each focused on a specific topic. The module examined in this study focused on GenAI in education. Students learned how GenAI tools function, including the design of large language models and their response-generation mechanisms, and engaged with resources on effective prompting and evaluating response accuracy. Figure 2 presents the outline of the GenAI module. The module culminated in a group discussion, with prior activities designed to prepare students for it.

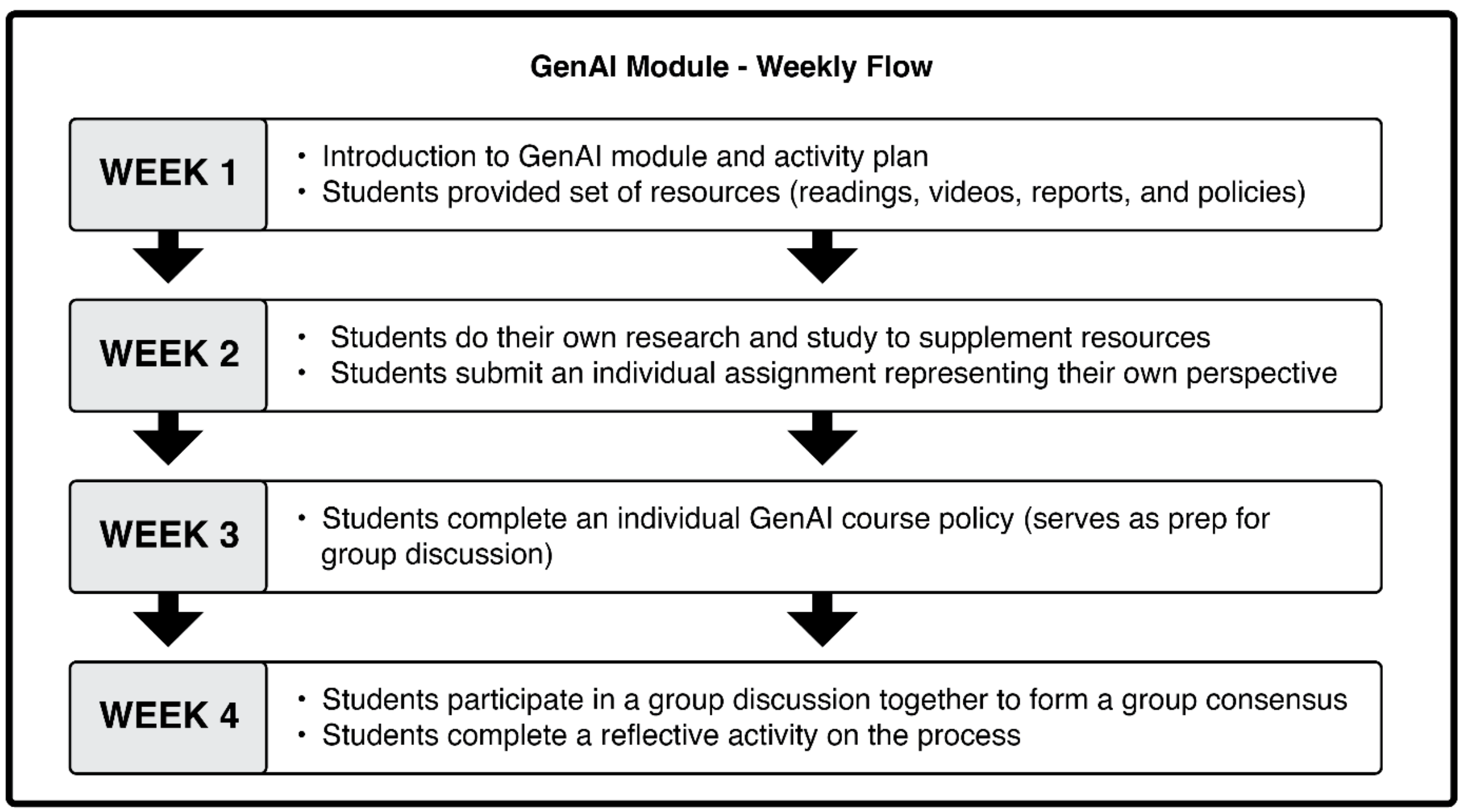

*Figure 2: Weekly flow of activities in the course, from introducing the topic to students to finally having them participate in the group discussion.*

The objective of this module was twofold: first, to provide students with an understanding of how GenAI functions; second, to explore its implications across various fields and tasks. Because the module assumed students would have access to GenAI tools, teaching them to use these technologies effectively was considered essential to supporting their academic work.

B. Resources (Readings, Videos, Reports, and Policies)

To help students jump-start their research process, we provided curated resources during the first week (presented in Table 1). These resources cover discussions on GenAI's technical and societal dimensions, aligning with the broader curriculum students encounter throughout the course. Students were encouraged to use these resources as a starting point and to search for other items to continue building their course policy. Each element of the course policy needed to be justified with evidence supporting its inclusion.

*Table 1: AI resources provided to students to help them navigate the module's content.*

| | Media Title | Media Type | Citation |
|---|---|---|---|
| 1 | AI Governance in Higher Education: Case Studies of Guidance at Big Ten Universities | Peer-Reviewed Article | (Wu et al., 2024) |
| 2 | The urgent risks of runaway AI - and what to do about them | Video (Ted Talk) | (Marcus, 2023) |
| 3 | How does artificial intelligence learn? | Video (Ted Talk) | (Brownell, 2021) |
| 4 | How AI could save (not destroy) education | Video (Ted Talk) | (Khan, 2023) |
| 5 | Towards social generative AI for education: theory, practices and ethics | Peer-Reviewed Article | (Sharples, 2023) |
| 6 | Generative AI and Use of College Data Policy | Institutional Policy | (*Generative AI Policy*, 2024) |
| 7 | Generative Artificial Intelligence in Higher Education: Evidence from an Analysis of Institutional Policies and Guidelines | Peer-Reviewed Article | (McDonald et al., 2025) |
| 8 | Vetting Generative AI Tools for Use in Schools | Policy Brief | (Sallay, 2024) |
| 9 | Higher education crisis: Academic misconduct with generative AI | Peer-Reviewed Article | (Song, 2024) |
| 10 | Example AI Syllabus Statements from University of Virginia Faculty AI Guides | Guide | (Bruff, 2024) |

C. Individual Course Policy Design Assignment

In the third week of the module, students were tasked with describing elements they believed should be included in a GenAI course policy. For this exercise, each student was required to detail at least five elements but could present as many as needed to make a coherent policy. They needed to consider several aspects for each recommended element: its importance within the policy framework, the benefits of incorporating it, and any potential concerns other stakeholders might have, such as instructors or future employers.

Students completed this activity individually, drawing upon the provided resources as well as their own research. While they could reference justifications from the resources provided during the first week of the course, they were also encouraged to expand their scope beyond these initial materials. Students had the opportunity to incorporate insights from their own educational experiences, reflecting on times when GenAI supplemented their learning and times when they encountered challenges using it. Engaging with such personal reflections was a crucial part of the task and helped convey why the policy needed to be thoughtful and well-structured. Feedback was provided to students after they submitted the individual writing assignment.

### D. Group Discussion & Writing Assignment

Following the individual assignments, participants engaged in a group discussion with 4-6 peers to develop a course policy on GenAI. This group discussion lasted approximately 45 minutes. Initially, each participant had the chance to present their views, outlining what they believed should be included and detailing the specifics of each component. Discussion then shifted to crafting a comprehensive policy incorporating all perspectives, with participants questioning one another's viewpoints and considering implementation: what the policy would cover, when it would apply, how it would affect different groups of students or courses, and any additional factors necessary for success. They were asked to collate these into a final GenAI policy, which they each submitted individually. Figure 3 presents an excerpt of a GenAI course policy designed by a group of participants.

After the discussion, participants reflected on moving from individual to collaborative policy creation, describing whether their perspectives had changed and any new ideas encountered. Reflection occurred through a brief debrief immediately after the discussion and a short written assignment submitted by the end of the day. The researchers reviewed and revised both individual and group writing prompts before implementation.

**Generative AI (GenAI) Course Policy**

**1. Academic Integrity in GenAI Use**

*Policy*: Students must use GenAI tools as supplementary aids rather than for completing assignments directly. Using GenAI for cheating, such as generating entire essays, coding tasks, or solving exam questions, is strictly prohibited.

*Importance*: This reinforces the principle that academic achievements should reflect a student's individual capabilities and understanding. Misuse hurts the integrity of academic credentials.

*Concerns*: Students may misuse GenAI to avoid learning, which leads to not enough subject mastery. Additionally, ethical questions arise about fairness if some students use GenAI improperly while others follow the rules.

*Benefits*: Encourages students to leverage AI responsibly to enhance learning, such as clarifying concepts or generating ideas. Proper use develops both critical thinking and adaptability to evolving technologies.

*Human Oversight*: Instructors should review assignments for signs of over-reliance on GenAI and discuss appropriate use during the semester. Plagiarism detection tools integrated with AI usage checks can provide additional protection.

**2. Citation of GenAI Usage**

*Policy*: Students should disclose GenAI use in assignments by citing the tool.

*Importance*: Transparency regarding AI-generated content is vital for maintaining academic honesty and acknowledging sources. This policy ensures that student's credit GenAI appropriately when used in their work.

*Concerns*: Students might underreport their GenAI use to avoid penalties, or over-reliance could lead to diminished critical thinking skills. Failure to cite AI contributions might lead to questions about originality and accountability. It could also obscure how much of the work was independently done versus AI-assisted.

B*enefits*: Citing GenAI use sets a precedent for acknowledging AI contributions, a skill that may become increasingly relevant in professional settings. This practice promotes integrity and helps instructors see students' understanding of course material versus reliance on AI tools.

*Human Oversight*: Instructors should verify citations against assignment content and provide feedback on whether usage aligns with policy guidelines. Reviewing citations can provide insight into how GenAI is being integrated into learning.

*Figure 3: A snippet from a group's designed GenAI course policy.*

# IV. Methodology

## A. Participants

Sixty-eight (68) undergraduate students studying information technology and cybersecurity enrolled in the course and engaged with the module as part of their regular coursework. Fifty-five (55) students in the course agreed to participate in this study; data from students who did not consent were removed during the initial stages of dataset creation and were not included in the analysis. Consent was obtained from all participants for data collection and analysis in accordance with IRB protocols (the host institution granted IRB approval). Consent to participate in the study did not affect students' assessment or grades in the course.

Most participants were junior and senior undergraduate students with varying prior knowledge and experience with AI and GenAI. In course discussions, participants described differing course policies, ranging from the permitted use of GenAI to complete bans. Before the module and data collection, students had completed several weeks of discussions on how AI works and its social impact.

B. Data Analysis

We used a hybrid inductive-deductive thematic analysis approach to explore the data and identify themes while keeping our theoretical framework in focus. This approach draws on the inductive familiarization and theme-generation guidelines described by Braun & Clarke (2006) and the codebook-based, coding-reliability approach described by Boyatzis (1998), given our use of a structured framework and interrater reliability measures in the later stages of analysis. This hybrid approach has been used in other studies to explore policy design (Chiu, 2024).

Our analytic process began inductively. First, the research team familiarized themselves with the data by inductively reading the artifacts, including the assignments and group discussions, and writing and collating recurring ideas, tensions, and patterns. During this stage, we did not initially force the data into predetermined framework categories. Instead, we annotated the artifacts and noted both patterns that appeared relevant to the ED-AI Lit Framework and ideas that extended beyond or complicated it.

After this initial inductive stage, we moved to a deductive, framework-driven phase of analysis. We mapped the inductively identified patterns and data segments onto the ED-AI Lit Framework, using it to organize, compare, and interpret the emerging codes. The research team discussed segments that did not clearly align with the framework to determine whether they represented meaningful emergent insights, refinements to existing categories, or data outside the scope of the analysis. This process allowed inductive findings to inform subsequent framework-driven coding rather than being prematurely excluded.

We then created initial themes and subthemes by comparing the inductive codes with the framework categories. Cohen's Kappa (.84) was used to measure interrater reliability between the two coders. Any discrepancies were discussed and resolved until all reviewers agreed on the coding. Finally, we reviewed the themes and formalized meaningful names for each theme. The final themes largely reflected the ED-AI Lit Framework and incorporated emergent insights identified during the inductive phase.

To minimize potential bias, and as part of the IRB protocols, one researcher not involved in teaching the course led the initial review of the collected data. Additionally, student grades were posted before any data analysis to prevent the researchers' evaluations from influencing course grading or the study's findings.

## V. Findings

In this section, we present descriptive findings from the analysis, illustrated with supporting excerpts, and then provide a detailed presentation of the subthemes.

### A. RQ1: What practical factors do students prioritize in their proposed GenAI use policies, and how do they justify these choices?

To answer this question, we examined how the activity guided participants through the research and design process, encouraging them to integrate AI literacy readings, additional research, and their own experiences into implementable policies. Our analysis was organized around the ED-AI Lit Framework: *Knowledge*, *Evaluation*, *Collaboration*, *Contextualization*, *Autonomy*, and *Ethics*. A final theme captured responsibilities that participants attributed to the institution.

Rather than focusing solely on the policy items themselves, we explored the underlying reasons and intentions behind their inclusion to provide a more nuanced understanding of the course policy elements. Through the themes, we uncovered participants' statements of intent, which revealed the reasoning behind their proposed course policies and underscored the importance of considering what was included and why it was deemed necessary.

Across the seven themes, participants did not treat the GenAI course policy as a set of isolated rules. Instead, they framed effective policy as an interconnected system in which knowledge, evaluation, collaboration, contextualization, autonomy, ethics, and institutional responsibility mutually shaped one another. Several themes functioned as prerequisites for others: for example, students' ability to act autonomously depended on their knowledge of GenAI systems and their capacity to evaluate AI outputs critically. At the same time, participants surfaced tensions between flexibility and standardization, support and enforcement, and individual responsibility and institutional obligation. The results below therefore describe each theme while also highlighting how themes intersected, reinforced one another, or created points of tension in participants' policy recommendations.

We present Table 2, which provides excerpts representative of each theme and subtheme, followed by further discussion of each theme.

*Table 2: Themes, subthemes, and excerpts from participants' designed GenAI course policy.*

| **Theme 1: Knowledge** | |
|---|---|
| *Subtheme* | *Supporting excerpts from the GenAI course policy* |
| 1.1 Task-related training for students | *"Training and education are important to ensure students possess proper level of technological literacy to effectively use and understand GenAI's limitations."* |
| 1.2 Task-related training for faculty and instructors | *"The training will provide professors with know how to use GenAI to its best potential in their course teachings. By having experience with GenAI, professors can be more confident in their decision of whether they should include GenAI in their teachings or not."* |
| **Theme 2: Evaluation** | |
| 2.1 Awareness of the limitations of AI | *"Students should understand the limitations of GenAI, such as its inability to reason, its reliance on existing data, and its lack of awareness of context or nuance. This awareness is important to prevent misuse or unrealistic expectations of AI capabilities."*<br>*"Faculty can help inform students whether the content that they are being taught is beyond or within the scope of GenAI."* |
| 2.2 Critical Evaluation of Outputs | *"Students should be taught to critically evaluate GenAI outputs, as these tools often generate content with inaccuracies or biases. If students take AI-generated responses at face value, they risk including flawed information in their work."* |
| 2.3 Student's reliance on AI | *"Students must understand how to utilize and properly integrate AI to assist with learning and not relying on it completely to complete an assignment. Over reliance on AI can hinder long-term development amongst students, especially in the field of software engineering."* |
| **Theme 3: Collaboration** | |
| 3.1 Teaching about prompting | *"Training students on how to properly use GenAI to minimize misuse due to lack of understanding. For example, providing in-class example prompts on how to use GenAI for coursework assignments or provide the prompts to the students to use for their coursework assignments."* |
| 3.2 Disclosure and citation | *"Students must disclose when and how they have used GenAI in their assignments. This promotes transparency and helps educators understand the role AI played in the work."* |
| **Theme 4: Contextualization** | |
| 4.1 Supportive rather than punitive system | *"It's important that students be encouraged to learn and be productive in an environment where consequences for their mistakes are present, but reduced, since it's not reasonable to expect that they'll begin the course as experts."* |
| 4.2 Encouraged and prohibited use | *"Allowable applications might include seeking clarification on complex topics, brainstorming project ideas, or debugging code."*<br>*"Prohibited activities, such as submitting AI-generated work as one's own or bypassing critical thinking, must be explicitly outlined to preserve integrity."* |
| 4.3 Course customization | *"Professors of each class or department must clarify what students can use AI for. For example, English classes can allow students to use AI grammar check. IT classes can deny students from using AI to write their project code. Be specific about what students can use for each class."* |
| **Theme 5: Autonomy** | |
| 5.1 Accountability | *"Accountability is important as it ensures that students are held responsible for how they choose to use reinforcing academic integrity and upholding a fair assessment when it comes to grading."* |
| 5.2 Understanding consequences of misuse | *"Consequences for the misuse of GenAI intentionally or repeatedly but unintentionally should align with the honor code. This policy doesn't exist to punish students, rather it protects the reputation of the students who uphold the Honor Code and who use GenAI ethically and responsibly."* |
| **Theme 6: Ethics** | |
| 6.1 Access to tools for all | *"Providing free or discounted access to these tools fosters a level playing field, particularly for underrepresented or economically disadvantaged students. Without access, some students may be left behind, unable to fully participate in GenAI-enabled learning."* |
| 6.2 Sustainability of AI | *"[Using GenAI] can add to pollution and harm the environment. That's why it's important to use GenAI responsibly. For example, instead of asking GenAI a hundred random questions, think about what the students really need help with and ask only those questions."* |
| **Theme 7: Institutional Responsibility** | |
| 7.1 Approved list of tools | *"There should be an approved GenAI tool available to the university's users, etc. This section outlines what is available for the faculty, staff and students, and that no other GenAI tool should be used for official university work."* |
| 7.2 Ensuring privacy and security | *"This section of the policy should state how data is collected to help AI. It should state what is being collected and how it is being collected."* |
| 7.3 Actively review and update policy | *"Continuous feedback and revision are important due to the ever-changing nature of technology and innovation as GenAI evolves. Due to this evolution, policy should evolve with it and reflect the new capabilities of GenAI to make sure that only current and pertinent information is included and retained."* |

### i. Theme 1: Knowledge

The first theme explored how a comprehensive GenAI course policy must create opportunities for students to learn how AI works. Participants framed knowledge as foundational to nearly every other policy concern. Without a shared understanding of GenAI's capabilities, limitations, and risks, policies on evaluation, disclosure, accountability, and misuse were difficult to interpret and apply consistently.

In Subtheme 1.1, participants called for institutions and instructors to provide accessible, task-related training for students so they can, at a minimum, be exposed to the tools' pros and cons. Participants emphasized the importance of making this training available to all new students and of customizing it to the field or course in which it is presented, suggesting that AI literacy should be treated as context-specific rather than generic.

Similarly, Subtheme 1.2 participants called for instructor training, describing past experiences and potential scenarios where instructors acted on a flawed understanding of GenAI. These findings are consistent with existing research highlighting AI's unequal distribution and use across fields, underscoring the need for targeted training and education for instructors to address these gaps (Dewan et al., 2025). One example that emerged repeatedly was the use of AI checkers, which indicate the likelihood of AI use. Currently, these systems can produce erroneous reports on AI use, leading to accusations of unreported use by students (Saha & Feizi, 2025). This concern links knowledge to later issues of autonomy and accountability, as students may be held responsible under policies enforced through tools or assumptions that instructors may not fully understand.

### ii. Theme 2: Evaluation

Theme 2 builds on Theme 1, emphasizing the importance of critical evaluation skills when working with AI technologies. Throughout our analysis, participants consistently emphasized critical evaluation as a crucial method for promoting fairness when using AI in education. All participants agreed that developing these skills would help students make informed decisions about using AI and mitigate potential biases or misuse, both in learning and beyond the classroom.

In Subtheme 2.1, participants recognized AI's reliance on training data, which can shape its reasoning and yield unreliable or biased outputs. They called for these limitations to be described in the policy, suggesting that misuse may stem not only from intentional misconduct but also from insufficient understanding of AI's limits.

In Subtheme 2.2, participants highlighted the importance of critically evaluating AI outputs. Many participants emphasized the need for students to be aware of and carefully consider the sources and biases of the training data used by AI tools, as well as the potential limitations of these tools. Participants further suggested that instructors should provide clear guidelines and expectations for AI tools and emphasize the importance of understanding the underlying concepts and principles.

Finally, Subtheme 2.3 explored how course policies should address the role of AI systems in assignments and assessments. Some advocated for blanket policies that prohibit or restrict the use of AI tools across the course. Others recommended a more nuanced approach in which instructors discuss AI's limitations and potential biases with students. This variation suggests that participants viewed evaluation not only as an individual student skill, but also as a policy decision about whether courses should respond to AI primarily through prevention or through critical engagement.

### iii. Theme 3: Collaboration

Theme 3 emphasized equipping students with essential communication and collaboration skills with AI systems and individuals. Building on Themes 1 and 2, participants framed collaboration as more than simply using AI tools; it required knowing how to engage with AI productively, evaluate its contributions, and remain transparent about its role in student work.

In Subtheme 3.1, most participants argued that prompting activities or exercises should be included to better engage with GenAI tools. Many participants also emphasized that instructors should be proactive in ensuring that students can use AI tools effectively and efficiently as collaborative partners, a sentiment that many educators share (Dewan et al., 2025). Many asked instructors to provide sample prompts that illustrate real-world applications of GenAI in their field, while acknowledging potential pitfalls. This suggests that participants viewed prompting not as a technical shortcut, but as a teachable practice connected to disciplinary learning.

In Subtheme 3.2, participants called for effective and standardized ways to disclose and cite AI use. Participants argued that transparency is key in ensuring that students can collaborate with the tools without using them uncritically or invisibly. This subtheme reveals a tension between AI as a legitimate collaborator and as an unacknowledged substitute for student work, with disclosure serving both accountability and the appropriate use of AI-assisted collaboration.

### iv. Theme 4: Contextualization

Participants emphasized the importance of encouraging students to use AI in real-world settings. This theme built on earlier themes by showing that knowledge, evaluation, and collaboration depend on context: what counts as appropriate AI use may vary by discipline, assignment type, and learning goal.

In Subtheme 4.1, many participants highlighted the importance of creating an environment that encourages responsible use of AI tools rather than punishing use. They shared experiences of 'honor code' systems penalizing students with little evidence of misuse, revealing a tension between enforcement and support and a preference for policies that guide rather than punish.

Subtheme 4.2 highlighted the importance of clearly defining specific uses for AI tools within a course. These uses could include brainstorming ideas, generating research questions, seeking clarification on complex concepts, or troubleshooting code. By making these uses explicit, participants can understand what is and is not allowed when using AI. This approach also aligns with the evaluation idea presented in Theme 2, as students must evaluate whether AI use is appropriate and beneficial for their learning in a given task.

Subtheme 4.3 emphasized the need for course policies to be tailored to specific disciplines and courses rather than relying on a single, one-size-fits-all approach. Participants suggested that instructors consider each course's unique characteristics and requirements when developing policies related to AI use. This variation points to a broader policy tension: students may benefit from consistency across courses, but meaningful AI guidance often requires local adaptation to disciplinary and assignment-specific contexts.

### v. Theme 5: Autonomy

Participants emphasized the importance of promoting student autonomy and self-determination in their interactions with AI through the course policy. Building on earlier themes, autonomy depended on students having the knowledge, evaluation skills, and contextual guidance needed to make informed decisions about AI use while maintaining their personal agency.

Subtheme 5.1 focused on accountability in AI-related decision-making. Participants stressed that, since AI outputs are evaluated as representative of students' knowledge and skills, students must take responsibility for how they use AI and for its consequences. This subtheme revealed a tension between being encouraged to use AI productively and remaining accountable for the final work submitted.

Subtheme 5.2 highlighted the importance of clearly outlining the consequences for AI misuse and due process procedures within the course policy. Many participants shared experiences in which students were not adequately informed about due process expectations or about their ability to defend themselves if they believed they had been unfairly accused. This lack of transparency and clarity led to fear and ambivalence among students, who often felt uncertain about whether their use of AI was acceptable and remained fearful even when it was. By providing clear guidance on the consequences of misuse, course policies can reduce uncertainty and foster a more positive, fair learning environment.

### vi. Theme 6: Ethics

Participants focused on ethical concerns, including fairness, accountability, transparency, and privacy. This theme extended earlier discussions by showing that responsible AI use is not only a matter of individual student behavior but also depends on whether students have equitable access to tools and whether those tools can be used responsibly within broader social and environmental constraints.

Subtheme 6.1 centers on unequal access to AI tools, where cost becomes a significant barrier. Many participants called for standardized or affordable tools built into the policy, drawing an analogy to textbook costs that could disadvantage students even when not required, revealing a tension between encouraging AI use and avoiding amplified inequity.

Subtheme 6.2 focused on AI's environmental impact. Several participants shared research on AI systems' energy demands, calling for environmentally conscious policies and for discouraging the use of GenAI for tasks that do not meaningfully contribute to learning. Together, these concerns suggest that ethical AI policy must consider not only whether AI use is permitted, but whether it is necessary, equitable, and educationally justified.

### vii. Theme 7: Institutional Responsibility

Participants demonstrated awareness of responsibilities across multiple levels: students, faculty, and the institution. This final, macro-level theme emphasized the institution's critical role in establishing and implementing GenAI course policies to support fair, consistent, and secure AI use across the university.

In Subtheme 7.1, participants discussed the need for an approved list of AI tools licensed or sanctioned by the institution. Participants argued that this would enable instructors to provide more targeted training and help streamline course materials, making it easier for students to access reliable resources. This subtheme also connects to Theme 6, as institutionally approved tools could help address unequal access while providing more consistent protections for student data.

In Subtheme 7.2, most participants agreed that the institution is responsible for safeguarding student data when using GenAI. This includes inputs into AI and how instructors use them, such as grading responses. Participants stressed that having an approved list of services would enable the institution to protect user data better and maintain confidentiality, while also addressing the challenge of balancing the convenience of AI tools with the institution's obligation to protect student privacy.

Finally, in Subtheme 7.3, participants unanimously pushed for the notion that any course policy should be reviewed and updated regularly to ensure it effectively achieves its intended goals. Students also emphasized the importance of having a say in developing and revising course policies, viewing this as essential for building trust between the institution, instructors, and students. This suggests that institutional responsibility is not only technical or administrative, but also participatory: effective policy requires ongoing revision and student involvement.

Together, the seven themes suggest that participants viewed GenAI policy as an interconnected system: knowledge and evaluation supported responsible use, collaboration and contextualization shaped integration; and autonomy, ethics, and institutional responsibility addressed fairness, access, privacy, and accountability. These connections also revealed key policy tensions: flexibility versus consistency, support versus enforcement, and student autonomy versus institutional responsibility.

## B. RQ2: How do participants reflect on the process of designing a GenAI policy?

We used guided inquiry to help students reflect on AI course policy development while considering multiple perspectives and drawing on their own experiences. This framework encouraged students to examine their roles as learners alongside instructors' responsibilities in creating fair and effective learning environments. By asking students to hold both perspectives at once, the activity supported problem-solving, stakeholder awareness, and policy design that balanced student needs with institutional responsibilities.

Participants found the activity engaging and described how it changed or expanded their perspectives. Participants' reflections varied: some focused on the instructor's role in supporting compliance with AI guidelines, others emphasized the connection between student and instructor AI literacy, and still others highlighted the close relationship between fairness and learning. A representative participant response captured this sentiment:

> *Preparing for the discussion really opened my eyes to the complexities of integrating GenAI into educational settings. One of the biggest takeaways for me was understanding the balance needed between supporting innovation and ensuring responsible use. Before this, I mostly thought of GenAI as a helpful tool for tasks like brainstorming or research. However, the discussion made me realize just how important it is to have clear policies that define its use. My perspective changed as I began to see GenAI not just as a convenience, but as something that requires careful oversight. Without guidelines, it could be misused or even lead to ethical challenges. This discussion helped me appreciate the importance of structure when using such a powerful tool, and I now understand how clear expectations benefit students and educators.*

Beyond this individual sensemaking, some participants' reflections focused more directly on the collaborative process, describing how engaging with peer and instructor perspectives shaped their thinking. One participant reflected how the guided inquiry process supported not only individual sensemaking but also perspective-taking across stakeholder roles:

*Hearing [peer]'s view shifted my view on environmental sustainability in relation to AI use. [Peer] did a good job explaining what they researched about AI energy use and it convinced me.*

Table 3 presents the topics participants explored in their reflections after the group discussion, offering further insight into their thoughts and concerns regarding AI integration. In addition, we present Figure 4, which displays the frequency of each topic. Figure 4 is intended as a descriptive supplement rather than an indicator of thematic significance. Frequency counts show where coded discussion clustered across thematic areas, but were not used to determine theme importance or inclusion. We present this data to illustrate the potential value of the activity as a learning experience. Although the frequency of a topic is not the most critical factor, many participants found the discussion helpful for gaining new information and perspectives. We anticipate that while some participants may have gained valuable insights or perspectives, others' personal views may remain unchanged. Guided inquiry, by design, supports individual sensemaking rather than convergence toward a single conclusion.

*Table 3: Topics and subtopics from participants' reflections on the module.*

| Topic | Subtopic |
|---|---|
| *Topic 1: Learned Something new about GenAI* | Environmental impact and sustainability |
| | Ethics and evaluation |
| *Topic 2: Acknowledge other stakeholders' perspectives* | Understanding a peer perspective |
| | Instructor perspective |
| *Topic 3: Acknowledge changes to personal perspective* | Balancing tradeoffs and benefits |

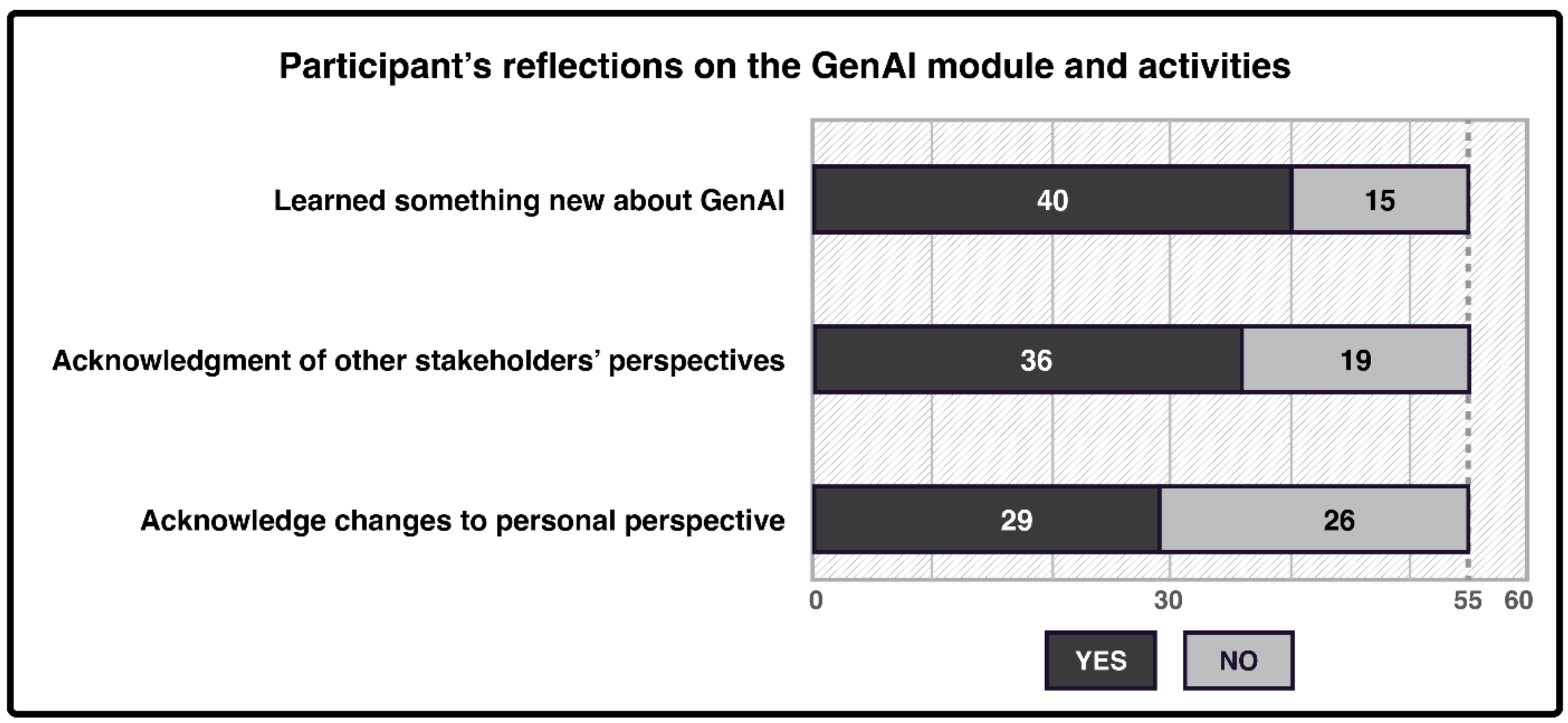

*Figure 4: Participant's reflections after completing the module, grouped by common themes.*

### i. Topic 1: Learned something new about GenAI

Most participants reported learning something new through the GenAI module. Participants were surprised to learn about the significant environmental impact of GenAI models, including the resources required to train them and the raw materials needed to produce the specialized hardware used in AI development. Similarly, many participants were unfamiliar with ongoing academic discussions about bias, fairness, and transparency in AI use.

### ii. Topic 2: Acknowledge other stakeholders' perspectives

About two-thirds of the participants acknowledged that the activity encouraged engagement with perspectives beyond their own, including those of their peers, instructors, institutions, and potential employers. This shift towards stakeholder-centric thinking affected participants' perceptions of GenAI's potential benefits and drawbacks. As they grappled with these perspectives, participants recognized the need for balanced approaches that consider the interests of multiple stakeholders, and many developed deeper empathy for instructors and institutions navigating the complexities of GenAI integration.

### i. Topic 3: Acknowledge changes to personal perspective

In addition, participants' personal views on GenAI's role in education changed after participating in the activity. Approximately half of the participants reported a shift in their perspective. A common thread among many participants was the growing awareness of the tradeoffs of using GenAI. By examining these benefits against the costs, many participants said they came to understand instructors' motivations and decision-making processes better, recognizing that instructors are not simply reacting to external pressures or trends, but navigating complex factors that shape their teaching methods and institutional goals. This growing awareness of trade-offs echoes the tensions participants identified in their policy recommendations (RQ1), suggesting that the same considerations that shaped their proposed policies also informed their personal understanding of the design process.

## VI. Discussion and Implications

Given this study's preliminary, context-bound scope, several key takeaways emerged. Firstly, participants overwhelmingly described the module positively. The guided inquiry approach structured their engagement through active exploration and meaning-making, consistent with Kuhlthau et al. (2007)'s model of constructed, rather than passively received, understanding. Participants' varied prior experiences with GenAI course policies reflect the broader landscape in which institutions and instructors have adopted widely divergent approaches, including prohibition, conditional use, and silence on the issue altogether (Ali et al., 2025; McDonald et al., 2025; Wang et al., 2024). The guided inquiry activity served both as a graded course assignment and as a source of research data, a dual role that may have influenced student responses. At the same time, this context reflects the authentic conditions under which policies are negotiated in instructional settings, offering insight into how students reason about GenAI use when evaluation and accountability are present.

Across their reflections and participation, students expressed a desire to be involved in shaping GenAI course policy, while also signaling uncertainty about how to contribute meaningfully. This tension mirrors longstanding concerns in the student voice literature about the gap between students being consulted and students exercising genuine influence over educational decisions (Cook-Sather, 2006). From a pedagogical design perspective, this suggests that co-design activities may benefit from additional scaffolding that supports students in moving from surface-level input toward what Hart (1992) describes as "substantive" participation, where students can genuinely articulate values, negotiate trade-offs, and understand their role in governance-oriented discussions.

Some students used the activity to pursue interests aligned with prior assumptions or anecdotal claims. Others found it reaffirmed their desire to explore AI tools further, a pattern educators may view with ambivalence, given their feelings of being "unprepared" (Dewan et al., 2025). This points to guided inquiry's potential to accommodate exploratory pathways and perspectival thinking within AI literacy contexts, while also building awareness of the complexity of policy design.

Students' discussions of the instructor's role further highlight design considerations for using AI in the classroom. Participants frequently linked student preparedness for AI-enhanced learning to instructors' AI literacy, a connection supported by frameworks that position educator AI literacy as foundational to effective instruction (Allen & Kendeou, 2024). This suggests that student-facing activities may be most effective when situated within broader instructional and institutional learning efforts, particularly given how much institutional policy contexts shape the conditions under which both instructors and students engage with GenAI (Dewan et al., 2025; McDonald et al., 2025). At the same time, participants raised questions about responsibility and authority in policy formation, underscoring the need for communication between instructor, administrative, and student roles. They also caution against arrangements in which student input is solicited but does not substantively shape outcomes (Cook-Sather, 2006; Rudduck & Fielding, 2006).

From a design perspective, the guided inquiry activity can be adapted across disciplines to support AI literacy while aligning with varied course contexts, extending beyond specific technical fields (Almatrafi et al., 2024). Because policies, rules, and standards are common across professional fields, using them as design topics can also help students consider the future of work in their disciplines, drawing on approaches that use real-world governance scenarios to develop capacity for responsible technology use (Hingle & Johri, 2024b). When adapting this activity, instructors should frame AI not only as a technical tool but as a sociotechnical system embedded in human practices, institutional structures, and ethical commitments, helping students examine questions of responsibility, acceptable use, and collective impact that policymakers actively navigate

Future studies could examine whether participation in GenAI policy co-design activities leads to changes in students' actual AI use behaviors. For example, researchers might compare students' AI use before and after the activity through reflective logs, assignment disclosure statements, surveys of AI use frequency and purpose, or analysis of how students evaluate and revise AI-generated outputs. Such work could help determine how policy co-design influences not only students' stated values and concerns, but also their everyday decisions for GenAI use.

Future studies could also focus on encouraging cross-disciplinary collaboration among students. Given that participants noted disciplinary differences in how GenAI was understood and applied, future work might also investigate cross-disciplinary implementations to support perspectival thinking and examine whether students' policy priorities shift when they engage with peers from other fields. Finally, research on how student-led policy discussions can translate into institutional decision-making would help bridge the gap between classroom discussions and broader policy development efforts.

# VII. Limitations

This study is exploratory, focusing on the potential of guided inquiry, perspectival thinking, and student involvement in co-designing AI course policies. Limitations of the activity's design suggest the findings are context-specific. We intend for them to inform future research and the iterative development of participatory AI literacy interventions.

This study and the module we designed were tailored specifically to our class participants, and the findings should be interpreted in that context. Because the study did not include a baseline or pre-intervention measure, the findings are not intended to demonstrate changes in students' AI literacy or learning gains attributable to the intervention; rather, the reported themes reflect how students articulated their perspectives, reasoning, and concerns in response to the guided inquiry activity, which may include pre-existing views. In designing the course, we drew on prior knowledge of participants' interests, particularly their interest in discussing GenAI, and their disciplinary background, which informed our expectation that the activity would resonate with this group. Consequently, replicating this activity in other contexts would require similar attention to participant characteristics, rapport, and local relevance.

# VIII. Acknowledgments

This work is partly supported by US NSF Awards 2319137, 1954556, and a USDA/NIFA Award 2021-67021-35329. Any opinions, findings, conclusions, or recommendations expressed in this material are those of the authors and do not necessarily reflect the views of the funding agencies.